\documentclass[preprint,floatfix,showpacs]{revtex4}
\usepackage{graphicx}

\usepackage{amsmath,amssymb}

\def\bea{\begin{eqnarray}}
\def\eea{\end{eqnarray}}
\def\beq{\begin{equation}}
\def\eeq{\end{equation}}

\begin{document}

\title{Nonlocal SU(3) chiral quark models at finite temperature: \\
the role of the Polyakov loop}

\author{Gustavo A. Contrera$^{a,b}$, Daniel G\'omez Dumm$^{c,d}$ and
Norberto N. Scoccola$^{a,d,e}$}

\affiliation{
$^a$ Physics Department, Comisi\'on Nacional de Energ\'{\i}a At\'omica, \\
 Av.\ Libertador 8250, (1429) Buenos Aires, Argentina.\\
$^b$ Universidad Nacional de San Mart{\'\i}n, Argentina.\\
$^c$ IFLP, CONICET $-$ Dpto.\ de F\'{\i}sica, Universidad Nacional
de La Plata, C.C. 67, (1900) La Plata, Argentina.\\
$^d$ CONICET, Rivadavia 1917, (1033) Buenos Aires, Argentina.\\
$^e$ Universidad Favaloro, Sol{\'\i}s 453, (1078) Buenos Aires, Argentina.}


\begin{abstract}
We analyze the role played by the Polyakov loop in the description of the
chiral phase transition within the framework of nonlocal SU(3) chiral
models with flavor mixing. We show that its presence provides a
substantial enhancement of the predicted critical temperature, bringing it
to a better agreement with the most recent results of lattice
calculations.

\end{abstract}


\maketitle

The detailed knowledge of the phase diagram for strongly
interacting matter has become an issue of great interest in recent
years, both from the theoretical and experimental points of view.
On the theoretical side, even if a significant progress has been
made on the development of ab initio calculations such as lattice
QCD~\cite{All03,Fod04,Kar03}, these are not yet able to provide a
full understanding of the QCD phase diagram due to the well-known
difficulties of dealing with finite chemical potentials. In this
situation, it is important to develop effective models that show
consistency with lattice results and can be extrapolated into
regions not accessible by lattice calculation techniques. In
previous works~\cite{GDS00,GDS02,GDS05,DGS04} the study of the
phase diagram of SU(2) chiral quark models that include nonlocal
interactions~\cite{Rip97} has been undertaken. These theories can
be viewed as nonlocal extensions of the widely studied
Nambu$-$Jona-Lasinio model~\cite{reports}. In fact, nonlocality
arises naturally in the context of several successful approaches
to low-energy quark dynamics as, for example, the instanton liquid
model~\cite{Schafer:1996wv} and the Schwinger-Dyson resummation
techniques~\cite{RW94}. Lattice QCD
calculations~\cite{Parappilly:2005ei} also indicate that quark
interactions should act over a certain range in momentum space.
Moreover, several studies~\cite{BB95,BGR02} have shown that
nonlocal chiral quark models provide a satisfactory description of
hadron properties at zero temperature and density. On the other
hand, when looking at the description of the chiral phase
transition, it has been noticed that for zero chemical potential
these SU(2) models lead to a rather low critical temperature
$T_{cr}^0$ in comparison with lattice results~\cite{GDS00,GDS02}.
The aim of the present work is to go one step beyond these
previous analyses, studying the finite temperature behavior of
nonlocal chiral models that include mixing with active strangeness
degrees of freedom, and taking care of the effect of gauge
interactions by coupling the quarks with the Polyakov loop. The
inclusion of the Polyakov loop has been considered recently in the
context of NJL-like models (so-called PNJL
models)~\cite{Meisinger:1995ih,Fukushima:2003fw,Megias:2004hj,
Ratti:2005jh,Roessner:2006xn}, serving as an order parameter for
the deconfinement transition . In particular, this has been done
in Ref.~\cite{Blaschke:2007np} in the framework of a nonlocal
two-flavor model, focusing on the analysis of mesonic
correlations. Here we show that within nonlocal SU(3) effective
models with flavor mixing the coupling to the Polyakov loop leads
to a significant increase of the chiral phase transition
temperature $T_{cr}^0$, which is otherwise as low as in SU(2)
symmetric models. As stated, this is a desired effect in order to
be in better agreement with lattice expectations~\cite{lattc}.

We deal here with the nonlocal covariant SU(3) quark model
described in Ref.~\cite{Scarpettini:2003fj}, including now the
coupling to the Polyakov loop. The Euclidean effective action for
the quark sector of this model is given by
\begin{eqnarray}
S_E &=& \int d^4x \ \left\{ \bar \psi (x) \left[ -i \gamma_\mu
\partial_\mu + \hat m \right] \psi(x) - \frac{G}{2} \left[
j_a^S(x) \ j_a^S(x) + j_a^P(x) \ j_a^P(x) \right] \right.
\nonumber \\ & & \qquad \qquad \left. - \frac{H}{4} \ T_{abc} \left[
j_a^S(x) j_b^S(x) j_c^S(x) - 3\ j_a^S(x) j_b^P(x) j_c^P(x) \right]
\right\}\;, \label{se}
\end{eqnarray}
where $\psi$ is a chiral $U(3)$ vector that includes the light quark
fields, $\psi \equiv (u\; d\; s)^T$, and $\hat m = {\rm diag}(m_u, m_d,
m_s)$ stands for the current quark mass matrix. For simplicity we consider
the isospin symmetry limit, in which $m_u = m_d=\bar m$. The currents
$j_a^{S,P}(x)$ are given by
\begin{eqnarray}
j_a^S (x) &=& \int d^4 y\ d^4 z \ r(y-x) \ r(x-z) \  \bar \psi(y)
\ \lambda_a \ \psi(z)\, , \\ j_a^P (x) &=& \int d^4 y\ d^4 z \
r(y-x) \ r(x-z) \  \bar \psi(y) \  i \gamma_5 \ \lambda_a \
\psi(z)\, ,
\end{eqnarray}
where the form factor $r(x-y)$ is local in momentum space, namely
\begin{equation}
r(x-y) = \int \frac{d^4p}{(2\pi)^4} \ e^{-i(x-y) p} \ r(p) \;,
\end{equation}
and the matrices $\lambda_a$, with $a=0,..,8$, are the usual eight
Gell-Mann $3\times 3$ matrices ---generators of SU(3)--- plus
$\lambda_0=\sqrt{2/3}\;\openone_{3\times 3}$. Finally, the
constants $T_{abc}$ in the t'Hooft term accounting for
flavor-mixing are defined by
\begin{equation}
T_{abc} = \frac{1}{3!} \ \epsilon_{ijk} \ \epsilon_{mnl} \
\left(\lambda_a\right)_{im} \left(\lambda_b\right)_{jn}
\left(\lambda_c\right)_{kl}\;.
\end{equation}

The coupling to the Polyakov loop can be implemented by assuming
that the quarks move in a background color gauge field $\phi = i
A_0 = i g\ \delta_{\mu 0}\ G^\mu_a \lambda^a/2$, where $G^\mu_a$
are the SU(3) color gauge fields. Then the traced Polyakov loop,
which is taken as order parameter of confinement, is given by
$\Phi=\frac{1}{3} {\rm Tr}\, \exp( i \phi)$. In what follows we
work in the so-called Polyakov gauge, in which the matrix $\phi$
is given a diagonal representation $\phi = \phi^3 \lambda^3 +
\phi^8 \lambda^8$, which leaves only two independent variables,
$\phi^3$ and $\phi^8$.

To proceed we consider the grand canonical thermodynamical
potential of the model within the mean field approximation. Using
the standard Matsubara formalism we get
\begin{eqnarray}
\Omega_{\rm MFA}(T) & \ = \ & -\, 2\, T \sum_{f,c}  \int
\frac{d^3p}{(2\pi)^3}\sum_{n=-\infty}^{\infty} \ \mbox{ Tr ln }\left[
\omega_{nc}^2 + \Sigma_{fc}^2(\omega_{nc}^2)\right] \nonumber \\
& &  -\; \frac{1}{2}\left[ \sum_f (\bar \sigma_f \ \bar S_f  +
\frac{G}{2} \ \bar S_f^2) \; + \; \frac{H}{2} \, \bar S_u\ \bar S_d\ \bar S_s
\right] \; + \; {\cal{U}}(\Phi ,T) \ , \label{ommfa}
\end{eqnarray}
where $f=(u,d,s)$, $c=(r,g,b)$, and we have used the definition
$\omega_{nc}^2 = (\omega_n - \phi_c)^2 + \vec p\ ^2$, $\omega_n =
(2n+1)\pi T$ being the usual Matsubara frequencies. The quantities
$\phi_c$ are defined by the relation $\phi = {\rm
diag}(\phi_r,\phi_g,\phi_b)$. The constituent masses $\Sigma_{fc}$
are here momentum-dependent quantities, given by
\begin{equation}
\Sigma_{fc}(\omega_{nc}^2) \ = \ m_f\, + \, \bar\sigma_f\,
r^2(\omega_{nc}^2) .
\end{equation}
Within the stationary phase approximation, the mean field values of the
auxiliary fields $\bar S_f$ turn out to be related with the mean field
values of the scalar fields $\bar \sigma_f$ by~\cite{Scarpettini:2003fj}
\begin{equation}
\bar \sigma_u + G\,\bar S_u + \frac{H}{2} \, \bar S_u \bar S_s = 0\ \ ,
\qquad
\bar \sigma_s + G\,\bar S_s + \frac{H}{2} \, \bar S_u^2 = 0\ \ .
\label{gapeq}
\end{equation}

The effective potential ${\cal{U}}(\Phi ,T)$, which accounts for the
Polyakov loop dynamics, can be fitted taking into account group theory
constraints together with lattice results, from which the temperature
dependence can be estimated. Following Ref.~\cite{Roessner:2006xn} we take
\begin{equation}
{\cal{U}}(\Phi ,T) = \left[-\,\frac{1}{2}\, a(T)\,\Phi^2 \;+\;b(T)\, \ln(1
- 6\, \Phi^2 + 8\, \Phi^3 - 3\, \Phi^4)\right] T^4 \ ,
\end{equation}
with the corresponding definitions of $a(T)$ and $b(T)$. Owing to the
charge conjugation properties of the QCD Lagrangian~\cite{Dumitru:2005ng},
the mean field value of the Polyakov loop field $\Phi$ is expected to be a
real quantity. Assuming that $\phi^3$ and $\phi^8$ are real-valued
fields~\cite{Roessner:2006xn}, this implies $\bar\phi^8 = 0$, $\bar\Phi =
[ 2 \cos(\bar\phi^3/T) + 1 ]/3$.

For finite current quark masses the quark contribution to $\Omega_{\rm
MFA}(T)$ turns out to be divergent. To regularize it we follow the same
prescription as in previous works~\cite{GDS05}. Namely, we subtract from
$\Omega_{\rm MFA}(T)$ the quark contribution in the absence of fermion
interactions, and then we add it in a regularized form, i.e.\ after the
subtraction of an infinite, $T$-independent contribution. From the
minimization of this regularized thermodynamical potential, it is possible
now to obtain a set of three coupled "gap" equations that determine the
mean field values $\bar\sigma_u$, $\bar\sigma_s$ and $\bar\phi^3$ at a
given temperature.

We are also interested in the estimation of chiral condensates, which are
given by the vacuum expectation values $\langle\bar u u\rangle =
\langle\bar d d\rangle$ and $\langle\bar s s\rangle$. As usual, they can
be obtained by varying $\Omega_{\rm MFA}$ with respect to the
corresponding current quark masses. The explicit regularized expression
for a quark condensate $\langle\bar ff\rangle$ reads
\begin{equation}
\langle \bar f f\rangle \ = \ 4 \sum_{c}  \int \frac{d^3p}{(2\pi)^3}
\left\{ - \, T \sum_{n=-\infty}^{\infty} \ \left[
\frac{\Sigma_{fc}(\omega_{nc}^2)}{\omega_{nc}^2 + \Sigma_{fc}^2(\omega_
{nc}^2)} \; - \; \frac{m_f}{\omega_{nc}^2 + m_f^2} \right] +
\frac{m_f}{E_f} \, (n_{fc}^+ + n_{fc}^-) \right\} \ ,
\end{equation}
where $n_{fc}^\pm = \{1+\exp [(E_f\pm i\phi_c)/T]\}^{-1}$, with
$E_f = \sqrt{\vec p\ ^2+m_f^2}$.

In what follows we analyze the chiral phase transition for a definite form
factor, taking into account the temperature dependence of effective
masses, chiral condensates and susceptibilities. For simplicity, we have
considered a Gaussian form factor
\begin{equation}
r(p^2) = \exp{\left(-p^2/2\Lambda^2\right)}\ ,
\end{equation}
where $\Lambda$ is a free parameter of the model, playing the role of an
ultraviolet cut-off momentum scale. This parameter, as well as quark
current masses and couplings in Eq.~(\ref{se}), can be chosen so as to
reproduce the empirical values of meson properties at $T=0$. We take into
account the analysis in Ref.~\cite{Scarpettini:2003fj}, where the
parameters are fixed to obtain the empirical values of meson masses
$m_\pi$, $m_K$ and $m_{\eta'}$, together with the pion decay constant
$f_\pi$. As shown in that work, in this way one can get a good
description of the light pseudoscalar meson phenomenology at zero
temperature. For definiteness we will work here with the parameter set GI in
Ref.~\cite{Scarpettini:2003fj}. Namely, we use
\begin{equation}
\bar m = 8.5 \ {\rm MeV}\ , \quad m_s = 223 \ {\rm MeV} \ , \quad
\Lambda = 709 \ {\rm MeV} \ ,  \quad G\Lambda^2 = 10.99 \ , \quad
H\Lambda^5 = - 295.3 \ .
\end{equation}

\begin{figure}[htb]
\includegraphics[width=0.8\textwidth]{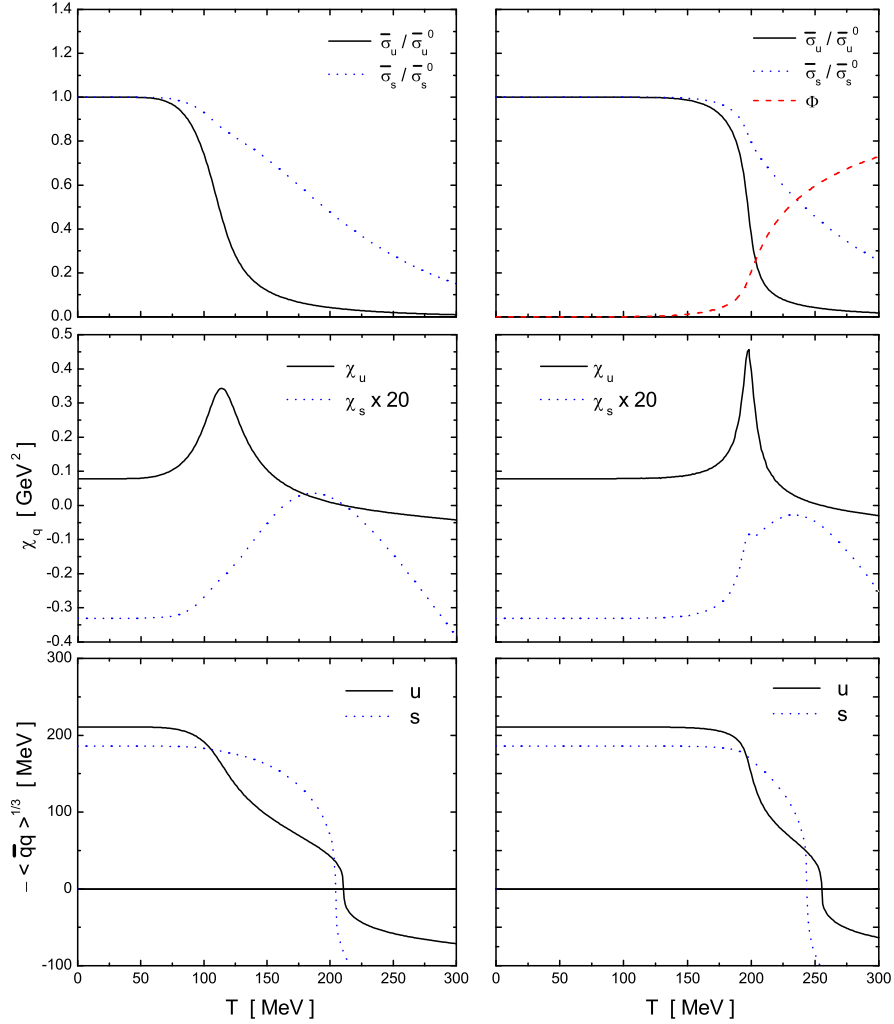}
\caption{Behavior of various physical magnitudes as functions of the
temperature. Left (right) panels correspond to the model in the absence
(presence) of a coupling to the Polyakov loop. The upper panels display
the behavior of the mean field values $\bar \sigma_q$ (normalized to their
value at $T=0$) and the Polyakov loop, while central and bottom panels
show the behavior of susceptibilities $\chi_q$ and chiral condensates,
respectively.}
\end{figure}

The corresponding numerical results are given in Fig.~1, where we show
the behavior of various physical magnitudes as functions of the
temperature within the described nonlocal SU(3) model. Left (right) panels
correspond to the model without (with) the inclusion of the Polyakov loop.
In the upper panels we show the behavior of the mean field values $\bar
\sigma_u$ and $\bar \sigma_s$ (solid and dotted lines, respectively). For
the sake of comparison, the curves have been normalized to the respective
values at $T=0$, which are found to be $\bar \sigma_u^0 = 304$~MeV and
$\bar \sigma_s^0= 427$~MeV (notice that at $T=0$ there is no effect of the
Polyakov loop). It is clearly seen that in both cases the SU(2) chiral
restoration proceeds as a smooth crossover, whereas there is an
enhancement of the order of 80~MeV in the corresponding critical
temperature when the effect of the Polyakov loop is taken into account. In
order to properly define the values of these critical temperatures, we
consider the chiral susceptibility $\chi_q$, defined as
\begin{equation}
\chi_q \ = \ \left(\frac{\partial\,\langle\bar qq\rangle}{\partial m_q}
\right)_{T={\rm cte}} \ .
\label{chiralsus}
\end{equation}
The phase transition temperature can be defined as the point where
the susceptibility shows a peak, the sharpness of this peak
serving as a measure of the steepness of the
crossover~\cite{GDS05}. The curves showing the behavior of the
susceptibilities in the nonlocal SU(3) models are plotted in the
central panels of Fig.~1. As already mentioned, it is seen that
the inclusion of the Polyakov loop leads to an enhancement in the
critical SU(2) chiral restoration temperature, which brings it to
a better agreement with the most recent lattice values
$T_{cr}^{0\,(Lat)} = 160 - 200$ MeV \cite{lattc}. Our results are
(see solid lines in the central panels of Fig.~1)
\begin{eqnarray}
\mbox{Nonlocal SU(3) model without Polyakov loop :} & & T_{cr}^0 \; = \;
114 \ {\rm MeV}
\nonumber \\
\mbox{Nonlocal SU(3) model including Polyakov loop :} & & T_{cr}^0 \; = \;
198 \ {\rm MeV} \nonumber
\end{eqnarray}
In addition, it is seen that after the inclusion of the Polyakov
loop there is an enhancement in the sharpness of the peak. The
restoration of the SU(2) chiral symmetry can be also observed when
looking to the behavior of the chiral condensates with the
temperature, which is shown in the lower panels of Fig.~1. At
$T=0$ we find $\langle \bar u u\rangle^{1/3} = - 211\ {\rm MeV}$
and $\langle \bar s s\rangle^{1/3} = - 186\ {\rm MeV}$. As usual,
the restoration of the SU(3) chiral symmetry is much less
pronounced, due to the larger current mass of the strange quark.
In fact, both in absence or presence of the Polyakov loop a broad
peak is observed in the corresponding chiral susceptibility
$\chi_s$. This is shown by the dotted lines in the central panels
of Fig.~1, where the values of $\chi_s$ have been normalized by a
factor 20 to be comparable with the corresponding values of
$\chi_u$. The peaks in $\chi_s$ are found to be placed at about
190 (230)~MeV for the model without (with) the Polyakov loop. It
is interesting to notice that due to flavor mixing effects the
light quark condensates $\langle \bar u u \rangle$ exhibit a
second drop at those temperatures. In addition, owing to flavor
mixing some signal of the SU(2) chiral restoration can be noticed
in the behavior of $\chi_s$: one can observe a shoulder at about
115 MeV for the model without Polyakov loop (barely noticeable in
the figure), and a small peak at about 200 MeV for the model with
Polyakov loop. Finally, it is also important to analyze the
behavior of the mean field value of the Polyakov loop field
$\bar\Phi$. As stated, the latter can be interpreted as an order
parameter of the deconfinement transition, $\bar\Phi =0$ and
$\bar\Phi = 1$ corresponding to confined and deconfined quarks,
respectively. In the nonlocal SU(3) theory under consideration,
the behavior of $\bar\Phi$ with the temperature is shown by the
dashed curve in the upper right panel of Fig.~1. By comparison
with the behavior of the mean-field value $\bar \sigma_u$, it can
be seen that even if the deconfinement transition goes smoother
than the SU(2) chiral restoration, both transitions are found to
occur at the same temperature region, located at about
$T=200$~MeV.

The analysis presented above corresponds to the specific case of a
covariant nonlocal SU(3) model, in which we have considered a separable
four-fermion interaction. Such an interaction is motivated by the
instanton liquid model~\cite{Schafer:1996wv}, where it arises from the
nonlocal structure of the QCD vacuum. For simplicity we have used here a
gaussian-like form factor, choosing a definite parameter set to get a good
agreement with meson phenomenology at zero temperature. In order to
determine the range of validity of the conclusions extracted from our
analysis, we have also considered different parameter sets (within the
range allowed by meson phenomenology), and different shapes for the form
factors such as e.g.~Lorentzian functions. We have also analyzed an
alternative way of including the nonlocality based on an effective
one-gluon exchange picture of strong interactions (see Ref.~\cite{Gomez
Dumm:2006vz} for a detailed comparison of model features in the SU(2) case
at $T=0$). It comes out that the results are qualitatively similar in all
these cases.

\hfill

To summarize, we have analyzed the role played by the Polyakov loop in the
description of the chiral phase transition within the framework of
nonlocal SU(3) chiral models with flavor mixing. We have found that its
presence provides a substantial enhancement of the predicted critical
temperature, bringing it to a better agreement with the most recent
results of lattice calculations. Another interesting effect of the
coupling to the Polyakov loop is that the phase transition becomes
steeper, showing a sharper peak in the chiral susceptibility. We notice
that, as it was done for the two-flavor case in
Ref.~\cite{Blaschke:2007np}, it would be of great interest to extend the
present SU(3) model analysis beyond the mean field approximation,
considering the effect of mesonic correlations. Other tasks to be
addressed are the analysis of the behavior of meson properties with the
temperature, and the extension of all these studies in the presence of
finite chemical potentials. We hope to report on these issues in
forthcoming publications.

\hfill

This work has been supported in part by CONICET and ANPCyT (Argentina),
under grants PIP 6009, PIP 6084 and PICT04-03-25374.

\end{document}